\documentclass[11pt]{article}

\title{The Genesis of Einstein's work on the problem of motion in General Relativity}
\author{Dennis Lehmkuhl}
\usepackage{epsfig, amsfonts, amsmath, amssymb, amsthm, pictex, verbatim, amsthm, pdfsync, mathrsfs}
\usepackage{tensind}
\usepackage[square]{natbib}
\usepackage{endnotes}
\usepackage[normalem]{ulem}

\begin{document}

\pagestyle{headings}

\bibliographystyle{agsm}
%\bibliographystyle{plainnat}

%\citationstyle{agsm} \harvardparenthesis{square}
%\harvardyearparenthesis{square}

\setcounter{secnumdepth}{4} \setcounter{tocdepth}{4}

\theoremstyle {plain} \theoremstyle{definition}
\newtheorem{definition}{Definition}[section]
\newenvironment{myquote}[1]{\begin{quote}\par\it}{\end{quote}}
\newcommand{\vs}{\vspace{3mm}}
\newcommand{\ci}[1]{#1 \hspace{2mm}\citet{#1}}
\newcommand{\conn}{\Gamma^\nu_{\phantom{\mu}\mu \sigma}}
\newcommand{\conna}{\Gamma^a_{\hspace{1,5mm}bc}}
\newcommand{\riem}{R_{\mu \nu \sigma}^{\quad \omega}}
\newcommand{\riema}{R_{abc}^{\quad d}}
\newcommand{\ric}{R_{\mu \nu}}
\newcommand{\ein}{G_{\mu \nu}}
\newcommand{\weyla}{C_{abcd}}
\newcommand{\weyl}{C^{\mu}_{\phantom{\mu} \nu \sigma \rho}}
\newcommand{\M}{M_{\mu \nu}}
\newcommand{\g}{g_{\mu \nu}}
\newcommand{\e}{\eta_{\mu \nu}}
\newcommand{\ga}{g_{ab}}
\newcommand{\gu}{g^{\mu \nu}}
\newcommand{\gd}{\sqrt{-g}}
\newcommand{\nn}{\nabla_{\nu}}
\newcommand{\nm}{\nabla_{\mu}}
\newcommand{\T}{T_{\mu \nu}}
\newcommand{\R}{R_{\mu \nu}}
\newcommand{\K}{K_{\mu \nu}}
\newcommand{\Q}{Q_{\mu}}
\newcommand{\G}{G_{\mu \nu}}
\newcommand{\F}{F_{\mu \nu}}
\newcommand{\LaM}{\mathscr{L}_M (g_{\mu \nu}, g_{\mu \nu, \sigma}, g_{\mu \nu, \sigma \rho})}
\newcommand{\LaG}{\mathscr{L}_G (g_{\mu \nu},\Psi,\Psi_{,\mu}, \Psi_{,\mu \nu})}
\newcommand{\ELG}{\frac{\delta \mathscr{L}_G}{\delta g_{\mu \nu}}}
\newcommand{\ELM}{\frac{\delta \mathscr{L}_M}{\delta g_{\mu \nu}}}
\newcommand{\la}{\lambda}
\newcommand{\tr}{(\g, \conna, \riema)}
\newcommand{\dotss}{\hspace{1mm} . \hspace{1mm} . \hspace{1mm} .
\hspace{1mm}} 
\newcommand{\ts}[1]{\textsc{#1}}
\newcommand{\n}{\noindent}
\newcommand{\A}{A_{\mu}}
\newcommand{\pst}{t_{\alpha}^{\tau}}
\newcommand{\gamman}{\gamma_{\mu \nu}}
\newcommand{\gammab}{\bar{\gamma}_{\mu \nu}}
\newcommand{\gammabb}{\bar{\bar{\gamma}}_{\mu \nu}}
\newcommand{\chris}{\genfrac{\{}{\}}{0pt}{}{\alpha \beta}{\nu}}

\thispagestyle{empty}

\begin{center}
\begin{Large}	
General Relativity as a Hybrid theory: The Genesis of Einstein's work on the problem of motion 
\end{Large}	
\end{center}

\begin{center}
Dennis Lehmkuhl, \\  Einstein Papers Project, California Institute of Technology \\ 
\smallskip Email: lehmkuhl@caltech.edu \\ \bigskip
Forthcoming in \emph{Studies in the History and Philosophy of Modern Physics} \\ Special Issue: Physical Relativity, 10 years on 
\end{center}

\begin{abstract}

In this paper I describe the genesis of Einstein's early work on the problem of motion in general relativity (GR): the question of whether the motion of matter subject to gravity can be derived directly from the Einstein field equations. In addressing this question, Einstein himself always preferred the vacuum approach to the problem: the attempt to derive geodesic motion of matter from the vacuum Einstein equations. The paper first investigates why Einstein was so skeptical of the energy-momentum tensor and its role in GR. Drawing on  hitherto unknown correspondence between Einstein and George Yuri Rainich, I then show step by step how his work on the vacuum approach came about, and how his quest for a unified field theory informed his interpretation of GR. I show that Einstein saw GR as a hybrid theory from very early on: fundamental and correct as far as gravity was concerned but phenomenological and effective in how it accounted for matter. As a result, Einstein saw energy-momentum tensors \emph{and} singularities in GR as placeholders for a theory of matter not yet delivered. The reason he preferred singularities was that he hoped that their mathematical treatment would give a hint as to the sought after theory of matter, a theory that would do justice to quantum features of matter.     	
	
\end{abstract}

\tableofcontents

\section{Introduction}
\label{S:Intro}
John Wheeler was a brilliant physicist, but his most outstanding talent lay in capturing the essence of physics in a picture, in a slogan. Generations of physicists and philosophers of physics have been struck by the simplicity and imagination-inducing power of Wheeler summarising the meaning of Einstein's field equations thus:\footnote{The quote is from \citet{wheeler1990journey}; \citet{MTW:1973}, p. 5, put it slightly differently. \citet{Wheeler-Ford:1998}, p. 236, give a bit of background story to the sentence.} 

\begin{quote}
Spacetime tells matter how to move; matter tells spacetime how to curve.
\end{quote}

Harvey Brown's brilliance, first and foremost, lies in putting his finger on a puzzle, and about not being seduced by a pretty picture of the physics. One of the most enduring series of questions of his 2005 book `Physical Relativity' is due to a raised eyebrow brought about by the fact that everybody else was so mesmerised by Wheeler's beautiful slogan. What does `spacetime' refer to in this slogan? \emph{How} does it tell matter how to move? Or does it?

Once we start pondering these questions, we quickly see that only the second part of the sentence is straightforwardly about the Einstein field equations. Yes, the Einstein equations can be interpreted as telling us that the presence of the mass-energy of matter (represented by $\T$) in a given region of spacetime increases the curvature in that region (represented by $\G := \R - \frac {1}{2} \g R$); i.e.

\begin{equation}
	\G = \kappa \T
	\label{E:full Einstein equations}
\end{equation}

But if $\T$ is the mass-energy of matter, how do we read `Spacetime tells matter how to move' into this? The answer is that the first part of Wheeler's sentence is best not interpreted as being directly about the Einstein field equations. It is about the fact that the Einstein equations \emph{imply} that (test) particles move on geodesics of the spacetime metric that is subject to the Einstein equations. It is this result that Brown had in mind when he wrote that\footnote{\cite{Brown:2007} p. 141 and 163, emphasis in original.} 

\begin{quote}
GR is the first in the long line of dynamical theories, based on the profound Aristotelian distinction between natural and forced motions of bodies, that \emph{explains} inertial motion. [...] For the first time since Aristotle introduced the fundamental distinction between natural and forced motions, inertial motion is part of the dynamics. It is no longer a miracle.
\end{quote}

Identifying geodesic motion with inertial motion, Brown describes how the Einstein equations imply that the Bianchi identities $\nabla^\mu \G \equiv 0$ imply that $\nabla^\mu \T = 0$, which in turn implies that test particles move on the geodesics of the affine connection compatible with the metric $\g$ governed by the Einstein equations. I have called this general approach of deriving the geodesic motion of matter `the T approach' in \citet{Lehmkuhl:2017a}. As \citet{malament2012remark} and \citet{Weatherall:2012brief} have pointed out, the T approach necessarily assumes that $\T$ obeys certain energy conditions.\footnote{See  \citet{Weatherall:2017ES} for a detailed analysis of the role of the energy conditions in the T approach, and a comparison between deriving geodesic motion in GR as compared to geometrized Newtonian theory. See \citet{Lehmkuhl:2017a} for a discussion of how the energy conditions needed in  the T approach square with the assumptions of the vacuum approach discussed below.}

Einstein wrote his first paper on the possible derivability of geodesic motion from the field equations in 1927, together with Jakob Grommer. Like Brown in the quote above, Einstein and Grommer emphasise how different this would make GR as compared to  previous field theories. 

However, Einstein and Grommer do not endorse the same approach of deriving geodesic motion. They explicitly consider the T approach and dismiss it in favour of not introducing any energy-momentum tensor whatsoever. Instead, they assume the vacuum field equations, $\R = 0$, and allow for singularities to appear in the spacetime metric in regions supposedly filled with matter.\footnote{We get $\R = 0$ from the full Einstein equations (\ref{E:full Einstein equations}) by setting $\T =0$, which implies that $T^\mu_\mu = 0$, which in turn implies that $R^\mu_\mu = R = 0$.  In short, the full Einstein equations imply that the scalar curvature $R$ vanishes for all vacuum solutions, and that the Einstein tensor reduces to the Ricci tensor. } This is all the more striking as Einstein had vigorously opposed singularities in debates with de Sitter and Weyl,\footnote{See \citet{EarmanEisenstaedt:1999} for details.} and had pioneered a restricted version of the T approach with Marcel Grossmann in 1913. What had changed between 1913 and 1927? And what can it teach us about GR -- about its history, but also about GR itself?

\section{The big picture: GR as a hybrid theory}
\label{S:bigpicture}
In a previous paper on the problem of motion in GR (\citet{Lehmkuhl:2017a}) I focused on the question of whether the Einstein-Grommer approach of 1927 is bound to represent matter by singularities in the spacetime metric; the claim that the approach is committed to this had been the main source of criticism in the philosophy of physics literature. I concluded that the Einstein-Grommer argument was independent of seeing particles as represented by singularities, and that the approach deserved more attention than it had attracted so far. However, in this previous paper I bracketed the question of how Einstein himself had seen the approach and its assumptions, and also what kind of perspective on GR had convinced him to follow this approach rather than the T approach. It is this question I want to address in the current paper. We will encounter a unique and, paradoxically, virtually unknown perspective on GR: Einstein's perspective. We will see how Einstein saw GR as a hybrid theory, both fundamental and effective. In particular, we will see how Einstein's approach to the problem of motion shows how he thought of GR as fundamental in its treatment of pure gravitational fields, but as effective/phenomenological in its treatment of matter and the interaction between matter and gravity. 

I do believe that reconstructing Einstein's path and his arguments for taking it is a worthy goal in its own right. But it is all the more fascinating that the resulting perspective on GR turns out to be useful for the modern interpretation of the theory, and I believe that even philosophers who are not interested in the history of the theory in its own right would  make a mistake to ignore it.

Work in the history of science often starts with a puzzle. Our puzzle, which I shall call Kennefick's puzzle, starts in 1913, when Albert Einstein and Marcel Grossmann worked on a generalisation of special relativity that was supposed to extend the relativity principle to accelerated motion while at the same time --- through the equivalence principle --- delivering a relativistic theory of gravity. They called their new theory an \emph{Entwurf} --- a draft, an attempt, a blueprint --- of a generalised theory of relativity.\footnote{See \citet{norton1995eliminative}, \citet{Rennjanssen:2006} and \citet{raz2015outline} for detailed analysis of the Entwurf theory.} All the guiding principles that would lead to Einstein's final theory of general relativity were already in place in the Entwurf theory, all the mathematical tools at hand, and even the structure of the gravitational field equations was similar (though more cumbersome and not generally covariant) to Einstein's field equations of late November 1915. Einstein and Grossmann even addressed the question of how the gravitational field equations are related to the equations of motion of particles subject to a gravitational field. They had already seen that the equivalence principle ought to be realised by taking particles subject only to gravity as moving on the geodesics of the spacetime metric governed by the gravitational field equations. What is most crucial for us is that already in 1913 Einstein had given an argument that for the case where the matter in question is a collection of pressureless particles (relativistic dust), the geodesic movement of said particles follows from the assumption that the covariant divergence of the energy-momentum tensor of the dust system vanishes. In other words, assuming that energy-momentum is covariantly conserved allows you to derive that the particles move on geodesics.\footnote{\citet{EinsteinGrossmann:1913}, sec. 4.}

% XXX It is definitely true that Einstein and Grossmann saw that nabla T = 0 implies the EOM. But did they also see that the derivative of the Entwurf gravitation tensor vanishes and that thus this follows from the field equations? Dan does not say this in his endnote, but he does say it in his small cue paper on p. 115. 

Now, in the course of working on Volume 7 of the Collected Papers of Albert Einstein, Daniel Kennefick discovered that there is a document that is most likely a draft to Einstein's famous Princeton lectures of 1921 in which Einstein had claimed that in the sucessor theory of the Entwurf theory, namely, General Relativity,\footnote{Vol. 7 of the Collected Papers of Albert Einstein (CPAE from now on), Doc. 63, p.453. The draft was written about four months before the manuscript of the Princeton lectures were finalized. For a detailed analysis of this document see \citet{kennefick2005einstein}.}

\begin{quote}
``[the field equation] already contains the divergence equation and with it the laws of motion of material points.''
\end{quote}

What brought about this extra step, the claim that the field equations imply that the covariant divergence of $\T$ vanishes, which in turn implies the laws of motion of material particles? Something had happened in the meantime, between 1913 and 1921. By 1919, Einstein had learned that once you have the Einstein equations and settle for $\G := \R - \frac {1}{2} \g R$  as the gravitation tensor on the left hand side of the field equations, the contracted Bianchi identities tell you that the covariant divergence of $\G$ vanishes identically, and thus the covariant divergence of the energy-momentum tensor must vanish also.  $\nabla^\mu \T = 0$ is a consequence of the Einstein field equations, we do not have to introduce it as an independent assumption.

Einstein must have seen that at least for the case where the energy-momentum tensor is that of relativistic dust, this meant that the geodesic motion of matter follows from the 1915 gravitational field equations, due to the latter implying energy-momentum conservation.\footnote{There has been some debate over whether $\nabla^\mu \T = 0$  really expresses energy-momentum conservation since in general it does not correspond to an integral conservation law. The latter is only possible in spacetimes with Killing symmetries, i.e. in very special cases of the overall solution space of the Einstein equations. See  \citet{Hoefer:2000}, \citet{Lam:2011} and \citet{Read:forth} for further discussion. For my purposes this does not make a difference though, and I succumb to the convenience of speaking of the equation as the conservation law of energy-momentum.} And yet, in his Princeton lectures of 1921, the draft of which contains exactly this claim, indeed in all his published lectures  and papers up until 1927, Einstein introduces the gravitational field equations and the geodesic equation as the equation of motion of particles subject to gravity as independent assumptions. Apart from this one sentence in the draft of the Princeton lectures, there is not so much as a hint between 1913 and 1927 that Einstein might have thought that the geodesic equation might be derivable from the gravitational field equations. And yet in 1927 he gives exactly such a derivation, and argues that this derivation makes general relativity different from any field theory that came before it: he argued that it is the first field theory in which the field equations and the equations of motion of matter do not have to be introduced as separate assumptions. Why did Einstein never mention this possibility between 1913 and 1927? Why did he make the claim of derivability in the draft to the Princeton letcures, only to fall back on the earlier claim that the field equations and the equations of motion of particles are separate assumptions in the published version? And what brought about the paper of 1927? These two questions are Kennefick's puzzle.

Work in philosophy of science also often starts with a puzzle. If one looks at the 1927 Einstein-Grommer paper on the problem of motion, one immediately wonders why the authors dismiss the T approach to deriving the equations of motion. Why did they instead opt for the route that takes the vacuum field equations as its starting point, a route that seemed to commit one to representing particles as singularities in the metric? Is such a commitment not much more problematic than making certain well-defined assumptions about $\T$? What is the argument for the vacuum approach that Einstein and Grommer endorse?

These questions, historical and philosophical, can only be answered together. In order to understand why Einstein and Grommer in 1927 dismissed the T approach, a route that Einstein and Grossmann had pioneered in 1913, we have to understand how Einstein's own interpretation of the role of the energy-momentum tensor in GR evolved up until 1927. In order to give a conceptual/philosophical analysis of Einstein and Grommer's alternative route of representing material particles by singularities in the metric field, we have to understand what kinds of representational capacities Einstein associated with singularities and energy-momentum tensors, respectively.

\section[The role of the energy-momentum tensor]{How Einstein saw the role of the energy-momentum tensor in GR}
\label{S:T}

% Start with marble wood quote and contra Geometrization, then ask what it could mean instead. Then the subsections below, then finish with quotes on phenomenological interpretation of T from earlier sources in note 6 of Doc 55.

One of the most oft-cited quotes of Einstein is his weighing of the respective virtues of the two sides of the Einstein field equations from \citet{Einstein:1936}, p. 370:

\begin{quote}
[General Relativity] is sufficient --- as far as we know --- for the representation of experiences of celestial mechanics. But it is similar to a building, one wing of which is made of fine marble (left side of the equation), but the other wing of which is built of low-grade wood (right side of equation). The phenomenological representation of matter is, in fact, only a crude substitute for a representation which would do justice to all known properties of matter.
\end{quote}

This quote has uniformly been interpreted as Einstein taking the left part of the equations (the Einstein tensor) as fine marble because of its pristine geometrical nature, whereas he supposedly saw the right side of the equations as low-grade wood because they lacked similar geometric significance.\footnote{Prominent examples are \citet{Vizgin:UFTbook} and  \citet{Goenner:2004}.} However, we now know that Einstein did not subscribe to the geometric interpretation of general relativity that says that the theory `geometrizes gravity' or `reduces gravity to spacetime curvature'. Indeed, Einstein actively opposed this interpretation.\footnote{See \citet{Lehmkuhl2014:geo} for details.} But then what was his complaint about the energy-momentum tensor in the paragraph cited above? The answer lies in the second sentence, which claims that energy-momentum tensors provide only a `phenomenological representation of matter'. In this section, we will see what Einstein really meant by this, that this line of thinking (and the particular choice of words) goes back at least as far as the early 1920s, and that hidden behind this choice of words is an interpretation of general relativity that will be strikingly foreign to many modern philosophers of physics; though, I believe, far less so to modern astrophysicists and mathematical relativists. 

\subsection{How $\T$ became the right-hand side of the new gravitational field equations}

% - Start with T in theories leading up to Entwurf theory using Genesis volumes. Motivate T from special relativity but also from comparison with Nordstroem's theory in which the trace of T is the source.
% - For the above see section 11 (p.455 in pdf) of Norton's Nordstroem paper in the genesis volumes, which points to section 7 of the Entwurf paper in which Einstein necessitates taking \T as a source by going over from a scalar gravitational potential (for which he argues only the trace of \T can be the source) to a ten-component gravitational potential.
% - In the above sources it becomes clear that Einstein believed (at least until the Vienna lecture) that only the full \T as source could do justice to the equality of inertial and gravitational mass, Laue's fundamental result about the stress tensor. See in detail his Nordstroem paper, which also gives such a detailed discussion of different material systems. 

We know that Einstein's first thought of combining the projects of extending the principle of relativity and formulating a relativistic theory of gravity occurred in 1907, when he conceived of the equivalence principle.\footnote{See \citet{Norton:1989b} for the full story and an analysis of  Einstein's version of the principle. See also \citet{janssen2012twins, janssen2008no} for how the equivalence principle relates to Einstein's other principles, the relativity principle in particular.} However, he only really started to work on the project with full steam in 1911, when he   formulated a series of gravitational field equations in which gravity was represented as a scalar potential, just as in Newton's theory. The scalar potential doubled as the speed of light, which thus became a dynamical variable. These theories are things of beauty; but Einstein ultimately abandoned the approach.\footnote{See \citet{norton1995eliminative} for details.} What is important to us is that, naturally, in a theory in which the gravitational potential is a scalar, the source term has to be a scalar too. In Newton's theory of gravity this role was fulfilled by the mass density scalar, but in a relativistic theory of gravity this would not do, as mass had turned out to be equivalent to energy and thus also to momentum.\footnote{The meaning of `equivalence' here is still subject to debate; see \cite{Lange:2001} and \cite{Flores:2005}.} Indeed, in 1912, in the midst of finally buckling down to work towards a relativistic theory of gravitation, Einstein claimed that the development of the energy-momentum tensor by Hermann Minkowski, Max Abraham, Max Planck and Max von Laue was ``the most important new advance in the theory of relativity''.\footnote{See Vol. 4, Doc. 1 CPAE, and \cite{Lehmkuhl2011}, section 2.1,  for some context.} Max von Laue really deserved most of the credit here: it was he who had developed a general theory of energy-momentum tensors of arbitrary stressed bodies and relativistic fluids, and included Minkowski's energy-momentum tensor for the electromagnetic field as a special case.\footnote{See \cite{Norton:1992}, section 9.2.}

In August 1912, Einstein moved back to Zurich, and started his collaboration with Marcel Grossmann that would lead to the 1913 \emph{Entwurf} theory, which contained all the core elements of general relativity with the exception of generally covariant field equations. Most crucially, in the Entwurf theory Einstein transitioned from working with a scalar  potential to a ten-component potential $\g$. In section 7 of his part of the \emph{Entwurf} paper, Einstein compares the two approaches; for even though his own scalar theory of gravity had been abandoned, Gunnar Nordstr\"om's scalar theory was still a contender for the correct relativistic theory of gravity. Indeed, in his Vienna lecture of 1913 Einstein made it clear that he regarded Nordstr\"om's theory as the only serious rival to his own 1913 theory. Einstein's argument in the Entwurf paper for why a ten-component gravitational potential should be preferred over a scalar gravitational potential draws on the question of whether the corresponding source terms can do justice to the equality of inertial and gravitational mass (often called the weak equivalence principle). For Einstein, this meant that the total energy of a material system must give us a measure of how strong it is as a gravitational source. He argued that the only way to do this in a scalar theory of gravity was to take the trace of the energy-momentum tensor, $T^\mu_\mu$, as the source of the gravitational field, but that even then the total energy of a material system (defined as a volume integral over $T^\mu_\mu$) can only play this role for closed static systems, and only in their rest frames.\footnote{For details on Einstein's reasoning  see \cite{Norton:1992}; and \citet{Giulini:2008} for criticism of Einstein's argument leading up to the conclusion that scalar theories of gravity are incompatible with the (weak) equivalence principle.} Einstein concluded by arguing that the only way to do complete justice to the equality of inertial and gravitational mass is to have the entire energy-momentum tensor as the source of gravity, rather than a contraction of $\T$. In turn, this means that one of the major reasons why Einstein proceeded from a scalar gravitational potential to a ten-component gravitational potential was that he had convinced himself that the ten-component energy-momentum tensor had to be the source of gravity.\footnote{Of course, the remaining two years until Einstein found the field equations of November 1915 were focused on finding the correct left-hand side of the field equations. For a heroic reconstruction of each of Einstein's steps on this long and windy path see \citet{Norton:1984}, \citet{Rennjanssen:2006},  \citet{RennSauer2006}.} 

\subsection{$\T$ as a place-holder for a theory of matter not yet delivered}

% - Discsuss T in Entwurf theory, Einstein's letters on Hilbert (Einstein to Lorentz, 13 November 1916 [Vol. 8A Doc. 276]) See also Genesis Volume 2. page 418 of the pdf, note 139.

The above story seems to cohere well with Einstein's Leibnizian/Machian commitment that the mass-energy distribution should uniquely determine the gravitational field, a commitment he held in a strong form until about the late 1910s or early 1920s, and in a weakened form until 1933.\footnote{It is only in \citet{Einstein:1918} that Einstein dubs this requirement `Mach's Principle'. For analysis see \citet{Hoefer:1995,Hoefer:1994}, \citet{norton1995mach}, \citet{Barbour:2007}, \citet{BrownLehmkuhl:2017}. The change in how Einstein used the 1918 version of Mach's principle has not been appreciated in the literature up to know; I will address this development in detail in \citet{Lehmkuhl:forthOUP}. Roughly speaking, Einstein originally believed that $\T$ should uniquely determine $\g$ for \emph{every} solution of the field equations. When he found that this principle is violated both for his 1915 and his 1917 equations (which introduced the cosmological constant), he started using the principle as a selection principle between `good' and `bad' cosmological solutions (compare his distinguishing between `good' and `bad' singularities, elaborated on in section \ref{S:EG}). Up until 1933, Einstein argued repeatedly that good cosmological solutions, i.e. solutions possibly corresponding to the actual universe, were those in which Mach's principle was fulfilled.} It may thus be surprising that even in the early phase Einstein insisted that while the right-hand side of the Einstein equations, $\T$, ought to uniquely determine the gravitational potential $\g$, the theory ought not put \emph{any} constraints on matter, and on what kinds of mass-energy distributions are allowed in turn. 

We see a first inkling of this already in the \emph{Entwurf} theory of 1913. When Einstein and Grossmann give a variational formulation of their gravitational field equations, they just use $\T$ as a primitive and unanalysed source term, rather than attempting a variational formulation of the matter laws as well. We see this more clearly in \cite{Einstein:1916b}, Einstein's definitive treatment of GR using a variational formulation. He writes in the introduction:

\begin{quote}
	My aim here is to present the fundamental connections as transparently and comprehensively as the principle of general relativity allows. In particular, specific assumptions about the constitution of
matter should be kept to a minimum, in contrast especially to Hilbert's presentation.
\end{quote}

A bit more than two weeks after submitting the paper for publication, Einstein sent a copy to H.A. Lorentz, the man who was, with Hilbert, the prime pioneer of using variational methods in the context of GR. Einstein writes about the paper:\footnote{Einstein to Lorentz, 13 November 1916 (Vol. 8, Doc. 276 CPAE).} 

\begin{quote}
I am sending you simultaneously with this letter a little paper in which I showed how I think the relation between the conservation laws and the relativity postulate should be construed. I made an effort to present the matter as succinctly as possible, free from all unnecessary trimmings. In particular, I wanted to show that the thought of general relativity does \emph{not} limit the manifold of possible choices for the matter [Lagrangian] to a higher degree than the postulate of special relativity, since the conservation laws are satisfied by any choice of [the matter Lagrangian]. The choice made by Hilbert thus appears to have no justification.\footnote{Einstein speaks of `the Hamiltonian function' when he means what is today referred to as the Lagrangian; I replaced the former by the latter for ease of reading.}
\end{quote}

Hilbert's choice for a matter Lagrangian that Einstein alludes to is the assumption (following Mie) that all matter could be accounted for by using a matter Lagrangian that only depended on the metric, the electromagnetic vector potential and its first order derivatives. 

In the letter to Lorentz quoted above Einstein held back; he was much more unrestrained when writing about the matter to his close friend Paul Ehrenfest a few months earlier:\footnote{Einstein to Ehrenfest, 24 May 1916 (Vol. 8, Doc. 220 CPAE).} 

\begin{quote}
I do not care for Hilbert's presentation. It is unnecessarily specialised regarding ``matter'', unnecessarily complicated, and
not honest (= Gaussian) in its structure (creating the impression of being superhuman by obfuscating one's methods).
\end{quote}

One might not think it possible, but Einstein was even more blunt when writing of his paper to Hilbert's \emph{Meistersch\"uler} Hermann Weyl, and directly after taking Weyl to terms for his own assumption of a specific matter Lagrangian:\footnote{Einstein to Weyl, 23 November 1916 (Vol. 8, Doc. 278 CPAE).}

\begin{quote}
Unfortunately, even a preliminary choice of a [Lagrangian] for matter is still quite involved, so I prefer to do this only for special cases. Thus, for instance, your matter (in the actual sense) is nothing but infinitely fine, electrically charged dust. The reason for this is that you did not furnish your matter with surface tension or cohesion. Neither electrons, nor atoms, nor macroscopic matter can be represented in this way. [...]

Hilbert's ansatz for matter seems childish to me, in the sense of a child innocent of the deceits of the outside world. [...] Either way, mixing the solid considerations originating from the relativity postulate with such bold, unfounded hypotheses  about the structure of the electron or matter cannot be sanctioned. 
\end{quote}

What stands out from the above are two things. First, from very early on, indeed even before his Machian ideals came under pressure, Einstein believed that the gravitational field equations should not put more constraints on what kind of energy-momentum tensor is allowed than special relativity does. In other words, the dynamics of the gravitational field ought not constrain what is allowed to be a source of these dynamics. I believe that Einstein was very aware of the fact  that the energy-momentum tensor is \emph{not} a representation of matter itself, but only of its energetic properties. And this was fine in the context of GR: the theory was supposed to be a theory of  gravity first and foremost, and in this context all one needs to know about matter is its mass-energy distribution. Einstein's 1936 claim that the energy-momentum tensor is only a phenomenological representation of matter means exactly that: the tensor does not tell us all there is to know about matter, but only what we need to know in the context of a pure theory of gravity. Indeed, Einstein used the same language already in 1922, and elaborated on it. In his Princeton lectures, \citet{einstein1922vier}, pp. 33-34, we find the following statements:

\begin{quote}
\emph{Phenomenological description of the energy tensor of matter. Hydrodynamical equations.} \\
We know today that matter is built up of electrically charged elementary particles, but we do not know the field laws which ground the constitution of these particles. Thus, when investigating mechanical systems, we are forced to make use of an inexact description of matter, which corresponds to that of classical mechanics. 
\end{quote}

These statements are in the `special relativity' section of the Princeton lectures; nothing is changed about them in the `general relativity' section. We find similar language in \cite{EinsteinGrommer:1923}, Einstein's first paper on a unified field theory, in this case Kaluza's theory:

\begin{quote}
However, besides the $\g$ Mr. Kaluza introduces an [energy-momentum] tensor to describe material currents. But it is clear that the introduction of such a tensor only serves to give a preliminary, merely phenomenological description of matter ... .  
\end{quote}

Crucially, even in the draft to the Princeton lectures in which Einstein \emph{did} explicitly state the possibility of deriving the equations of motion of particles from the field equations via the T approach (see section \ref{S:bigpicture}), right before pointing to this possibility Einstein states 

\begin{quote}
Energ[y] tensor of matter and el[ectrodynamics] must be known from elsewhere (Maxw. hydrodynamics. (sic.)
\end{quote}

I believe this is part of the reason why Einstein was unhappy with the role of the energy-momentum tensor in GR; so unhappy in fact that when he moved from the draft to the final version of the Princeton lectures he preferred to introduce the equations of motion as a separate assumption, rather than deriving them through the T approach. For already in the draft to the Princeton lectures when he noted the possibility of deriving the equations of motion within the T approach, Einstein pointed out that this would only work if $\T$ was delivered by another theory like hydrodynamics. For Einstein, $\T$ was not really a part of GR as such --- it was a kind of `docking station' for other theories, theories of matter like hydrodynamics or electrodynamics, that would have an  energy-momentum tensor as a consequence of their dynamical equations. This was the first reason that made Einstein unhappy with the T approach. 

But still: what did he mean with the claim that these theories would produce only a `phenomenogical representation of matter'? What was he missing in these theories of matter, like classical fluid dynamics, that would deliver a $\T$ to be plugged into the right-hand side of the Einstein equations? 

In one word: he missed quanta. More precisely, Einstein believed that every accurate (in contrast to phenomenolgical) theory of matter needed to give an account of the quantum nature of matter. This should \emph{not} be misunderstood as referring to matrix and wave mechanics, not to mention Hilbert spaces and probability amplitudes. For Einstein, the quantum nature of matter that any fundamental theory eventually had to account for was about the observed properties of the then known elementary particles, i.e. the electron and the proton, and arguably the photon. In particular, a theory that would give an account of the quantum nature of matter would account for the discretization of masses, energy levels and of electric charge. In other words, the theory needed to account for the fact that there are no particles with an electric charge smaller than that of an electron, and that elementary particles come with very specific rest masses and electric charges. Given these facts, classical electrodynamics and hydrodynamics, both of which are continuum theories, could not be seen as an accurate description of matter, according to Einstein.

  The sources quoted above are already suggestive in this regard, but it becomes most clear in a letter Einstein wrote to his old friend Michele Besso, complaining again about the merely phenomenological qualities of $\T$ but also elaborating on what would have to be done to go beyond it:\footnote{Einstein to Michele Besso, 11 August 1926 (Vol. 15, Doc. 348 CPAE).}

\begin{quote}
But it is questionable whether the equation $R_{ik} - \frac{1}{2} g_{ik} R = T_{ik}$ has any reality left within it in the face of quanta. I vigorously doubt it. In contrast, the left-hand side of the equation surely contains a deeper truth.
\end{quote}

This, finally, gives us the key to unlock the meaning of the famous 1936 quote on marble and wood. Einstein seeing the left-hand side of the Einstein equations as fine marble and the right-hand side as low-grade wood has nothing to do with geometry. It is about quanta. He believed that the left-hand side of the Einstein equations gave an accurate picture of the gravitational field, but that the right-hand side of the equations did not give an accurate picture of matter, for it does not account for the quantum features of matter. It is only a docking station for results of theories like classical hydrodynamics and electrodynamics, which do not do justice to the quantum nature of matter either. Thus, $\T$ in GR is a place-holder for a theory of matter not yet delivered. This is the second, more important reason for why Einstein was unhappy with the role $\T$ played in GR. 

What to do then? Einstein saw two possible ways to remedy the situation, in order to go beyond a pure theory of gravity and develop a theory that is accurate both about gravity and about matter. He considered the first option in the Princeton lectures, shortly before complaining about the phenomelogical nature of $\T$. When discussing the way classical electrodynamics models electrons, he writes:\footnote{\cite{einstein1922vier}, pp.32-33. The second  and third sentence quoted are a footnote to the first sentence.}

\begin{quote}
If we can build upon Maxwell's equations at all, the energy tensor of the electromagnetic field is known only outside the charged particles. It has been attempted to remedy this lack of knowledge by considering the charged particles as proper singularities. But in my opinion this means giving up a real understanding of the structure of matter. It seems to me much better to admit our present inability rather than to be satisfied by a pseudo-solution (\emph{Scheinl\"osung}).
\end{quote}

In the years to follow Einstein turned to the second possible solution that is only hinted at here: a real understanding of matter, instead of the `pseudo-solution' offered by modeling material particles as singularities. Starting with \cite{EinsteinGrommer:1923}, Einstein tried to find unified field equations for both gravity and matter that would allow for solutions capable of representing both the interior and the exterior of quantum particles as continous fields. Einstein expected that the discretization of electric charge, as well as that of the mass-energy of electrons and light quanta, would appear as a consequence of overdetermined solutions. This would potentially allow one to derive quantum conditions, and the rest masses and charges of elementary particles, from a continuous field theory.\footnote{Another reason to be unhappy with the energy-momentum tensor in GR are the results by \cite{Tupper:1981,Tupper:1983a,Tupper:1983b}. He showed that knowing the energy-momentum tensor of a material system does not suffice to tell us what kind of matter is present. For example, one and the same mass-energy-momentum distribution $\T$ featuring on the right-hand side of the Einstein equations, and solving the Einstein equations for the same metric, can correspond either to an electromagnetic field or a viscous fluid. Knowing the energy-momentum tensor is just not sufficient to know which of these two material systems it is that interacts with the metric field.} 

Einstein laid out this research programme most clearly in \citet{Einstein:1923Bietet}. In the next section we will see that  hitherto unknown correspondence between Einstein and the mathematical physicist George Yuri Rainich holds the key as to how the vacuum approach defended in the Einstein-Grommer paper came about. We will see Einstein reconsider the alleged pseudo-solution of treating elementary particles as singularities in the field. Einstein is just as skeptical about the representational powers of singularities as about those of energy-momentum tensors, but  step by step he starts to suspect that maybe a pseudo-solution can be a stepping stone to the real solution of understanding of matter.

\section{The Einstein-Rainich correspondence}
 \label{S:Rainich}
The story starts with the 50th anniversary of H.A. Lorentz' doctorate on 11 December 1925. Lorentz, at this time 72 years old, still reigned supreme, at least in Einstein's mind, as the physicist worthy of more admiration than any other alive.\footnote{See \citet{kox1993einstein} for  the most insightful analysis of Einstein's and Lorentz' relationship.} And indeed, Lorentz was not rusty: just two months before the anniversary celebrations he had written Einstein a 10-page letter with  a new solution to Einstein's most recent attempt at a unified field theory, consistency checks of the theory as a whole, and corrections to the equations in Einstein's published paper.\footnote{See \citet{Einstein1925t}, and H.A. Lorentz to Einstein, 18 October 1925 (Vol. 15 Doc. 90) for Lorentz' letter, and Einstein to Lorentz, 22 October 1925 (Vol. 15 Doc. 94) for Einstein's answer.} The theory was the last paper in a long series of papers on what Einstein called `the Weyl-Eddington approach' at a unified field theory, starting from Weyl's and Levi-Civita's discovery that the affine connection $\conn$ can be defined independently of the metric $\g$, and Eddington's idea of basing a unified field theory only on the affine connection. Indeed, the 1925 paper that Lorentz comments on is the first paper that partly breaks with this idea of Eddington's, and instead introduces what is today called a mixed geometry: a theory in which the metric and the connection are introduced as independent variables, and where the connection is not assumed to be symmetric.%\footnote{XXX Palatini.}

Alas, Einstein quickly became  disillusioned with the approach, though he would come back to it decades later.\footnote{See \cite{Goenner:2004}, section 6.1 and \citet{Goenner2014}, section 7, for details on this approach in particular, and \citet{Vizgin:UFTbook}, \citet{vanDongen2010einstein} and \citet{Sauer:2015} for Einstein's unified field theory programme more generally.} As in his previous attempts at finding satisfactory field equations for the affine connection, one major problem for Einstein was that he could not find any solutions that could be interpreted as representing electrons. As mentioned in the previous section, Einstein clearly expected of such a solution that it could do justice to the fact that electrons only ever come with a specific electric charge and a specific mass; ideally, the solution would predict the charge and rest mass of the electron. 

As his contribution for the Lorentz festschrift a month later, Einstein would submit a little-known paper in which he aims to prove a meta-theorem about a whole class of theories, and effectively argues that they all are empirically inadequate when it comes to such particle solutions. The paper takes as its aim the proof of the following theorem:\footnote{\citet{Einstein1925xx07}, pp.330-331.}

\begin{quote}
	If it is true that the electromagnetic field ought to be represented by an antisymmetric tensor ($f_{\mu \nu}$) of rank 2, then it is impossible for there to be generally covariant equations which 1) have the negative electron as a solution, 2) do not have a solution corresponding to a positive electron of the same mass.
\end{quote}	

If true, then this theorem applies to  Einstein-Maxwell theory, to all of Einstein's theories based on an affine connection, and his most recent unified field theoru (UFT) based on a mixed geometry (plus, to Weyl's UFT and the original UFTs of Kaluza and Klein).\footnote{Note that the theorem crucially depends on which mathematical object represents which physical field, and on what counts as a `solution corresponding to a [positive or negative] electron'. Thus, the theorem is more about a class of theories with particular bridge rules, rather than a theorem that is about the inherent mathematical structure of these theories.} Einstein implies that thus all these theories are empirically inadequate, for he thought he knew (as did everybody else at the time) that the `positive electron' was the proton, and had been shown to be 2000 times heavier than the `negative electron'.\footnote{\citet{Treder:1975} argues that by virtue of this theorem Einstein should be credited with having predicted antimatter, and the positron in particular. I would agree that on the basis of this theorem Einstein \emph{could have} predicted the existence of the positron. But given that he is adamant in his belief that such a particle does not exist, and takes its non-existence  together with the theorem as an argument against all these theories, it seems a bit of a stretch to say that he predicted the positron.}

It is at this point that a new actor appears on the stage: a young Russian mathematician who had only just escaped prosecution in Russia and obtained a postdoc at John Hopkins University.\footnote{Rainich's given name was Yuri Germanovich Rabinovich, which he changed to George Yuri Rainich upon arrival in the USA. In \citet{Nichols:2013}, section 1.8, Rainich's daughter Alice claims that Rainich had been imprisoned by the communist government because he had been teaching relativity theory, which was deemed incompatible with Marxism. She also describes how, with the help of his wife and a prison guard who had been his student, Rainich managed to escape the prison and the country in late 1922, and how they flew via Istanbul to the United States, where they arrived in early 1923.} George Yuri Rainich was supposed to translate Einstein's  \citeyear{Einstei1910d} article on the opalescence of homogenous fluids and liquid mixtures near the critical state; the translation was to be published in a new volume edited by Jerome Alexander. Rainich did not only translate the paper but wrote a list of remarks to Einstein that pointed out typos, weaknesses in the argument and ways to improve it. As far as we know, Einstein never acknowledged receipt of these remarks forwarded to him by Alexander, but soon after Rainich wrote to Einstein directly. He wrote (somewhat cockily) that it may have escaped Einstein's attention but that he (Rainich) had shown that the Einstein-Maxwell equations already do everything one could ask of a unified field theory.\footnote{Rainich  had shown that under certain conditions the Ricci tensor determines the Faraday tensor up to an integration constant. Thus, the Maxwell equations could be written in terms of combinations of the Ricci tensor and its derivatives; the resulting version of Einstein-Maxwell theory is of fourth order in the metric. \citet{misner1957classical} would later speak of this approach as ``GR as an already unified field theory'', after Misner independently discovered some of Rainich's results. Note that the resulting theory is not really Einstein-Maxwell theory, as it introduces extra conditions on the Ricci tensor and  contains higher-order derivatives. For a bit more detail and analysis on whether this approach constitutes a `geometrization' of the electromagnetic field see \citet{Lehmkuhl2016:Super-Substantivalism}.} Einstein and Rainich exchanged some letters about their respective expectations of a unified field theory, and it becomes clear again (compare section \ref{S:T}) that for Einstein unified field equations of gravity and electromagnetism are only acceptable if they give quantum particle solutions.\footnote{ Instead of getting on board with Rainich's programme,  Einstein used Rainich's discoveries about the curvature tensor to give a new interpretation to the trace-free field equations he had suggested in \cite{Einstein:1919}, in a paper he finished in January 1926 (\citet{Einstein:1927xx19}).} 

After a few months of silence in their correspondence,\footnote{For a comprehensive summary of the correspondence between Einstein and Rainich see the Introduction to Vol. 15 CPAE, pp. xlii-l.} Rainich then wrote a short paper  criticising Einstein's interpretation of the meta-theorem of the Lorentz festschrift, and sent it for publication to Adriaan Fokker on the same day that he sent a copy to Einstein.\footnote{Both Einstein's paper and Rainich's critique appeared in the journal \emph{Physica Nederlandsch Tijdschrift voor Natuurkunde}, which was edited by Adriaan D. Fokker at the time. Fokker had been Lorentz' PhD student and research assistant, and would become Lorentz' successor  as director of research of the Teylers Museum in Haarlem, Netherlands, in 1928.} 

In this paper, Rainich claims that Einstein's meta-theorem does not imply a contradiction with experience. He argues that it would only do so if the Einstein field equations were linear, for only then would the existence of a solution representing a `negative electron' at rest and a solution representing a  `positive electron' at rest imply the existence of a solution which represents  both particles at rest with respect to one another. Rainich argues that \emph{this} would contradict experience because we know that in reality two such particles would move towards one another. But, he points out, in a non-linear theory like Einstein-Maxwell theory the non-linearity of the field equations implies that in general the superposition of two solutions to the field equations is not itself a solution, and thus we cannot be sure that a static two-body solution exists. 

It is noteworthy that both Einstein and Rainich work with puzzling notions of empirical adequacy here. Einstein's paper suggests that the existence of positive electron solutions  renders a theory empirically inadequate. This presupposes that any solution to the field equations needs to be related to the real world, and the existence of a solution be interpreted as a prediction of a physical system corresponding to that solution. However, in the past Einstein had more than once helped himself to the possibility of deeming certain solutions as `merely mathematical', not corresponding to the actual universe.\footnote{Einstein had just entertained the distinction between mathematical solutions and those representing (aspects of) the real universe. In \citet{Einstein1923g}, Einstein's retraction of his earlier claim that Alexander Friedmann had made a mistake in his calculations giving a non-static  cosmological solution to the Einstein equations, the last sentence in the draft reads: ``It turns out that next to the static solutions the field equations allow for dynamic (i.e., changing with the time coordinate) central-symmetric solutions for the structure of space; however, these solutions are hardly physically meaningful.'' Einstein deleted the last part of the sentence before submitting the remark for publication.} One might have the same opinion here: as long as the field equations also allow for a two-body solution that represents the actual motion of charged electrons exerting forces on one another, one might not be too bothered by a (non-representational) static two-body solution.\footnote{Indeed, in section \ref{S:EG} I show that using this distinction was crucial for the 1927 paper by Einstein and Grommer.} 

Be that as it may, in his answer a bit more than two weeks later, Einstein reiterated an argument from the addendum to his paper for the Lorentz festschrift in detail; again he demanded that a satisfactory theory ought to explain why a particle with positive electric charge and the same mass as the electron does not exist. As in his previous correspondence with Rainich, Einstein did not really address the contents of Rainich's arguments, but reiterated his own. 

This is when Rainich must have decided to be frank; it must have seemed to him that Einstein was under the impression that he (Rainich) did not understand his (Einstein's) arguments. But of course, not agreeing is not the same as not understanding. Thus, in his answer of 5 April 1926, Rainich wrote: `I believe that I have now completely understood your view, and now take the liberty to try to make my own clear to you.'' He continues:\footnote{Vol. 15, Doc. 245 CPAE. }

\begin{quote}
It seems to me that the problem you indicate can be traced back to your goal of characterizing the properties of particles by the ``field inside matter''. Obviously, that way you rob yourself of the possibility of studying the relations between two particles directly and are obliged to try to find these relations by comparison of the findings stemming from studying the respective particles independently. In my op[inion], it is more advantageous to trace the properties of particles back to the makeup of the field surrounding them; then it becomes possible to derive the relations between two particles from the properties of their shared field.
\end{quote}	

Rainich thus critized Einstein's programme of finding solutions to the field equations that would characterize both their exterior and their interior by continuous fields. Instead, he believed that we should characterize the properties of particles by deducing them from their exterior field. 

In what follows, Rainich gives a detailed account of how this research programme could be carried out, and emphasises again how crucial of a difference it makes whether the theory in question has non-linear or linear field equations. Among other things, Rainich described the notion of a `residue', which he had introduced in \citet{Rainich:1925c}. According to Rainich, a residue is a constant obtained by integrating a field over a three-surface surrounding a material particle. In his letter to Einstein, he argued that the mass and electric charge of a body could be obtained as residues.\footnote{Note that Weyl had given a similar construction for obtaining the mass of a body embedded in an asymptotically flat spacetime  in the 1922 5th edition of his book \emph{Raum-Zeit-Materie}; in his annotation to the 7th edition of the book,  J\"urgen Ehlers argued that Weyl foreshadowed the concept of the ADM mass introduced by Arnowitt, Deser and Misner in \citeyear{ADM}.} He goes on to claim that in a linear field theory no relationships between the residues can be deduced from the field equations; in contrast to a non-linear field theory like Einstein-Maxwell theory. 

Now Rainich really had Einstein's attention, though the latter was still skeptical. In his answer from 18 April 1926 (Vol. 15, Doc. 258), Einstein writes:  

\begin{quote}
I hurry to answer your letter, happy that you struggle with the same questions as I myself have for such a long time, to no avail. The cardinal question is of course whether one should think of electricity as continuous or made up of singularities. 
\end{quote}	

Note that these are the same two options, the same cardinal question, that Einstein had envisaged in his Princeton lectures of 1922, discussed above in section \ref{S:T}. Einstein then had taken the former option, noting that considering particles as singularities was but a pseudo-solution. And indeed, Rainich allowed for the particles that he characterized by the field surrounding them to correspond to singularities in the field. Einstein goes on:

\begin{quote}
The latter option seems easier at first sight, since one could then just stick with the Maxwell equations without adding anything to them. [...] I am convinced that one could find a strict solution on the basis of the gravitational equations + Maxwell equations, which would represent the case of two electrons at rest (as singularities). For the case in which the particles in question have no electric charge this has already been shown by Weyl and Levi-Civita (special case of axial symmetry).  This would prove that your plan cannot be carried out.	
\end{quote}	

The last two sentences are particularly important. Einstein effectively says: Fair enough, in a non-linear theory it is not a given that just because static one-body solutions exist  static two-body solutions exist as well. But Weyl and Levi-Civita have shown that \emph{they do}, at least for the case where the bodies are not charged and subject only to gravity. 

Einstein may have expected that this would be the end of it --- who can argue with an explicitly found exact solution to the field equations? But Rainich did not falter. With his ever balanced mixture of deference in the beginning and perseverance in the main body of his letters to Einstein he writes:\footnote{George Yuri Rainich to Einstein, 23 May 1926 (Vol. 15, Doc. 293 CPAE)} 

\begin{quote}
I cannot tell you how grateful I am for your letters, which give me the feeling that I am not working in a vacuum. - But I have to say that your last letter did not convince me that there is no hope to solve the fundamental problems with a field theoretic point of view.
\end{quote}

Of course Einstein also considered himself as taking a field theoretic point of view. But Rainich's was different: his field theoretic viewpoint involved focusing on the exterior fields of particles, rather than  characterizing their interior in terms of fields as well.
 
Rainich then explicitly cites the part of Einstein's letter that gives a reason why his (Rainich's) programme is allegedly hopeless. He answers: 

\begin{quote}
To this I would like to reply that if it is possible to find a solution to a set of field equations that has two electrons at rest, then this could prove that the field equations are unsatisfactory.

%Darauf möchte ich erwidern dass wenn es möglich ist für ein System von Feldgleichungen eine Lösung mit zwei ruhenden Elektronen zu finden es beweisen könnte dass dieses System unzulänglich ist.
\end{quote}

Of course, in his previous letter to Rainich Einstein had claimed that Weyl and Levi-Civita had found such a static two-body solution to the Einstein equations (rather than to the Einstein-Maxwell equations), which is subject to the same charge: massive particles ought to move towards one another under the influence of gravity, not remain at rest with respect to one another. 

It is between this letter by Rainich and Einstein's next letter that a fundamental change in Einstein's thinking takes place. In his letter from 6 June 1926 (Vol. 15 Doc. 300), Einstein writes: 

\begin{quote}
I completely agree with you on the main point. If a theory gives a solution which represents two electrons at rest, then it is inadequate. This was indeed the reason why I thought that I had to reject a theory which \emph{regards} electrons \emph{as singularities}. For I had thought to have seen that any such theory would have solutions with electrons at rest. But it now seems that I was wrong about this. Either way, this is the core question: a theory is sensible only if it allows to derive the equations of motion of particles without any extra assumptions. Whether or not the electrons are treated as singularities or not does not really matter \emph{in principle}.
\end{quote}

Note the change in the `cardinal question'. In his letter of 18 April the cardinal question was whether one should think of electrons as made up of continuous fields or of singularities. Now, 2 months later, it does not matter whether one treats electrons as singularities; the new  `core question' is whether a given (field) theory can derive the equations of motion of the particle in question without any further assumption (beyond the field equations).\footnote{In \citet{Einstein:1923Bietet}, p. 361, Einstein had expressed the hope that finding a solution to unified field equations that overdetermine a static, spherically symmetric solution such that it is capable of representing electrons, as described in section 3.2 above, might also determine the equations of motion of the electrons. In the following we shall see that the correspondence with Rainich inspired him to approach the problem from the other end: see whether the field equations of GR determine the equations of motion of particles, in the hope of getting an inkling of how the theory can be generalized to a theory that allows for
solutions capable of representing electrons.} Even more importantly, Einstein now believed that he had been wrong to think that a theory that \emph{does} treat particles as singularities would allow for a static two-particle solution. But had he not previously claimed that Weyl and Levi-Civita had found \emph{exactly} this kind of solution to the Einstein equations? What happened between Rainich's last letter of 23 May and this letter by Einstein of 6 June?

I conjecture that Einstein went back to the papers by Weyl and Levi-Civita that he had cited in his letter of 18 April.\footnote{It would be more cautious to ask: what happened between Einstein's last letter of 18 April and this letter of his from 6 June? After all, it could be that Einstein went back to the papers by Weyl without having been prompted by Rainich's answer to his letter of 18 April. But the more likely scenario is that Einstein was prompted by Rainich (in his letter of 23 May) not acknowledging the reference to Weyl and Levi-Civita as proof of the existence of a two-body solution.} For if he had, he would have seen that they are not in any straightforward way giving static two-body solutions. In the following section, we will see how Einstein and Grommer implicitly draw on Einstein's correspondence with Rainich in the introduction to their 1927 paper, in how they justify that in GR representing matter by singularities is preferable to representing it by help of energy-momentum tensors. They then motivate the vacuum approach to the problem of motion by appealing to a discussion of the Weyl solutions to the vacuum field equations. The following section of their paper then discusses these solutions in some detail, but Einstein and Grommer's interpretation differs markedly from Weyl's. Again using Einstein's letters to Rainich in the interpretation of the paper, I will argue that Einstein and Grommer effectively reinterpret the mathematical Weyl solutions as a theorem proving the \emph{non}-existence of \emph{physical} static two-body solutions. We will also see that this reinterpretation is strongly informed by Einstein's view of GR as giving an adequate picture of regions containing only gravitational fields, and as a merely phenomenological/effective theory when it comes to regions containing matter.  
\section{Good and Bad Singularities in the 1927 Einstein-Grommer paper} 
 \label{S:EG}
% Absolutely crucial: Vol. 8A, Doc. 286, Einstein's first reaction to the Reissner-Nordstroem solution, i.e. one representing a charged particle/black hole. First discussion of electrons as singularities and maybe even relevance of t equilibrium.
% Bring Besso sentence at the very end of the  section, going full circle back to end of previous section.

\subsection{The introduction to the 1927 Einstein-Grommer paper}

\noindent A. \emph{What follows from the linearity of field equations} \bigskip

The Einstein-Grommer paper starts by pointing out that in a Newtonian theory of gravitation and in Maxwell-Lorentz electrodynamics the field equations and the equations of motion of particles subject to the field are logically independent. To show this for electrodynamics, where the field equations in question are the Maxwell equations and the equation of motion for particles is the Lorentz force law, Einstein and Grommer immediately discuss the case of two electrons at rest.  

They assume  $\phi_1 = \frac{\epsilon_1}{r_1}$ to be the electrostatic potential of a single electron at rest, and the separate potential $\phi_2 = \frac{\epsilon_2}{r_2}$ to represent a different electron at rest. Now, they note that if $\phi_1$ and $\phi_2$ both solve the Maxwell equations separately, then $\phi_1 + \phi_2$ also solves the Maxwell equations, and we have a solution representing two electrons at rest with respect to one another. \citet{Rainich:1926b} had started in the same way. 

Now remember that in his last letter to Rainich, after much persistence by the latter, Einstein had finally written: ``I completely agree with you on the main point. If a theory gives a solution which represents two electrons at rest, then it is inadequate.'' We can read Einstein and Grommer as making exactly this point in the case of classical electrodynamics: the Maxwell equations by themselves are inadequate, for they allow a solution that represents two electrons at rest with respect to one another. But Einstein and Grommer now give a new twist to this thought and take it as a positive argument for the claim that the equation of motion of electrons, the Lorentz force law, is logically independent from the Maxwell equations, and needs to be postulated separately.\footnote{\citet{havas1989early}, p. 239, argues that Einstein and Grommer  overstate their case. Havas points out that even though a linear field theory cannot describe interactions between particles  by a solution of superposed one-particle states, it is possible for linear field equations to constrain the motion of a particle subject to the field.}

In line with \citet{Rainich:1926b}, but without citing him, Einstein and Grommer attribute the logical independence of field equations and equations of motion to the linearity of the field equations. This immediately brings them (and us) to the question of how the field equations and the equations of motion are related in a non-linear theory such as GR. Of course, Einstein was well aware that in all his presentations of the theory up to then he \emph{had} postulated the Einstein equations as gravitational field equations and the geodesic equation as the equation of motion of particles subject to gravity independently from each other. But now he asked, for the first time publicly: did he have to? 

\bigskip
\noindent B. \emph{Three perspectives on GR} 
\bigskip

Einstein and Grommer discuss three perspectives (\emph{Betrachtungsweisen}) of how the field equations and the equations of motion \emph{might} be related to one another. According to what they call the Newtonian perspective, the gravitational field equations are $\R = 0$, and the equation of motion of particles is the geodesic equation; both are postulated separately from each other.\footnote{It is  striking that Einstein and Grommer claim that \emph{this} perspective corresponds to Newtonian theory. After all, in the gravitational Poisson equation the mass density is the source of the gravitational potential, and it also occurs in Newton's second law, the equation of motion.  In that sense Newtonian theory might best be associated with Einstein's second perspective, as described below. However, it remains true that in both Newtonian theory and in this way of relating non-linear gravitational field equations and  equations of motion, both sets of equations have to be postulated separately.}

The second perspective is what I have called `the T approach' in \citet{Lehmkuhl:2017a}. Einstein and Grommer describe it as amending the field equations by the energy tensor $\T$ of matter and of the electromagnetic field, and write the resulting equations in the form:%\footnote{XXX Affine note, see endnote.}

\begin{equation}
	(\mathfrak{R}^{k}_{i} - \frac{1}{2} \delta_{k}^i \mathfrak{R}) + \mathfrak{T}^{k}_i = 0
\end{equation}

Einstein and Grommer immediately note that the field equations thus formulated demand input from outside of GR: 

\begin{quote}
The theory requires an amendment that cannot be obtained by the relativity principle alone: the $\mathfrak{T}^{k}_i$'s must be expressed by some sort of (continous) field quantities, and the differential equations determining them must be given. Only then would we have a complete theory. 
\end{quote}

Here we see the same attitude of Einstein's as described in section \ref{S:T}: the energy-momentum tensor is but a docking station or a place-holder for a different theory that offers field equations for the fields of which $\T$ describes only its energetic properties. 

The authors then note that even in the absence of such a matter field theory, it is not possible in GR to choose the $\T$ freely, for the Bianchi identities imply that $\T$ has to obey the condition $\nabla^\mu \T = 0$.\footnote{Einstein had not been aware of the Bianchi identities when he first introduced the final field equations of general relativity in 1915, but used them in deriving the conservation of energy-momentum of matter in \citet{Einstein:1919}. See \citet{Rowe:2002} for historical discussion.} It is of course exactly this condition that carries the derivation of the equations of motion in the T approach.\footnote{However, it needs to be noted that in a Lagrangian formulation the matter field equations also imply the conservation condition; see \citet{HawkingEllis:1973}, p. 64-67. For a detailed analysis of the conservation condition, how it and its role in the T approach should be interpreted, see \citet{Weatherall:2017a}, in this issue.} Einstein and Grommer note:  

\begin{quote}
If one assumes that matter is arranged along narrow ``world tubes,'' then by an elementary consideration one obtains from this the statement that the axes of those ``world tubes'' are geodesics (in the absence of electromagnetic fields). This means: the law of motion results as a consequence of the field law.
\end{quote}

As we saw in section \ref{S:T}, this is exactly the version of the T approach that Einstein and Grossmann had pioneered in 1913. This brings us back to Kennefick's puzzle: what did Einstein dislike about this route? Einstein and Grommer again point out that the T approach would only overcome the dualism between field equations and equations of motion for material particles if we had indeed found the correct matter field equations producing $\T$. With the discussion of sections \ref{S:bigpicture} and \ref{S:T} in the back of our minds we can see why Einstein discounts  classical electrodynamics and hydrodynamics for this job: they, and with them the energy-momentum tensors they produce, only give us a `phenomenological picture of matter', for they do not allow us to account for the quantum nature of elementary particles, meaning the discretization of mass, energy, and electric charge. 

This is what motivates Einstein and Grommer's third perspective, which they adopt for the rest of the paper.\footnote{In \citet{Lehmkuhl:2017a}, I have called this route a special case of the vacuum approach to the problem of motion in GR.} It assumes the existence of only gravitational fields, and allows for the gravitational field to become singular the closer one gets to elementary particles.\footnote{Strictly speaking Einstein and Grommer write that they assume the existence of only gravitational and electromagnetic fields, but that in what follows they will neglect the existence of the latter.} The aim is then to show that the purely gravitational field equations

\begin{equation}
	\R = 0
\end{equation}

allow one to derive the equations of motion of particles without any further assumptions. 

We may recall how much Einstein had opposed representing material particles by singularities in his Princeton lectures of 1922 (see section \ref{S:T}) and still continued to do so in his correspondence with Rainich in 1926. In the latter, he had insisted on what amounts to working on an amendment of the second perspective, one that would allow to account for the quantum nature of elementary particles. The final paragraph of the introduction to the Einstein-Grommer paper helps us reconstruct what he must have realised between Rainich's letter of 23 May and his own letter of 6 June 1926.

\bigskip
\noindent C.  \emph{The difference between linear and non-linear field equations}
\bigskip

In the final paragraph of the introduction, Einstein and Grommer claim that they had long wondered whether an equation of motion for the singularities could be derived from the gravitational field equations.\footnote{Note again that Einstein and Grommer's approach does not actually depend on representing particles by singularities; they might as well speak of the equations of motion of material particles which are not explicitly described themselves but characterized by their exterior fields (compare Rainich's letter from 5 April 1926 in quoted in section \ref{S:Rainich}).  I work this possibility out further in \citet{Lehmkuhl:2017a}. } However, they write that what discouraged them from taking this path was that in practice the linearized Einstein field equations approximate the real world rather well. Furthermore, they expected that starting from a solution to the linearised field equations one could, through successive approximation, get closer and closer to a solution of the exact equations. 

Of course, it is exactly this conviction that had come under fire in \citet{Rainich:1926b} and the ensuing correspondence between Einstein and Rainich discussed in section \ref{S:Rainich}. There the discussion was about the different structure of the solution space of linear field equations vs. that of non-linear field equations: while the former allows for superposition of two solutions to create a new one, the latter generally does not. In particular, Rainich had argued that while a linear theory would allow a static two-body solution to be built from the superposition of two static one-body solutions, the same is not true in a non-linear field theory.

While Rainich had compared different theories, Einstein and Grommer now apply this thought to the relationship between the non-linear Einstein equations and their own linearized limit. They argue that the difference between non-linear and linear field equations means that one cannot be sure that every solution to the linearized field equations approximates a solution to the non-linear field equations. In  particular, one cannot be sure whether the fact that there is a static two-body solution to the linearized gravitational equations means that there is a corresponding solution to the non-linear equations. 

Of course, in his letter to Rainich of 18 April 1926, Einstein had asserted the existence of exactly such a solution to the non-linear Einstein equations, and credited Weyl and Levi-Civita with having found it. Seven weeks later, in his letter of 6 June, Einstein backtracked on the claim that such a solution exists. This is all the more striking as the second section of the Einstein-Grommer paper is all about the --- indeed --- static and axially symmetric solutions that Weyl had found and that Levi-Civita and Bach had elaborated on. 

As we will see in the following, what Einstein realized was that even though he had originally thought that there is a solution in the Weyl class that represents a static two-body solution to the vacuum Einstein equations, when one looks at the details one can interpret Weyl's results as a non-existence proof of a physical two-body solution, if one allows certain types of singularities while disallowing others. Einstein was driven to do exactly that by his view of how matter features in GR.

\subsection{The Weyl class of solutions and the two-body solution to the Einstein equations}

When Einstein originally attempted to describe the orbit of Mercury around the Sun, he was sure that he would have to be satisfied with an approximate solution to the linearized vacuum Einstein equations to describe the exterior gravitational field of the Sun that Mercury was subject to.\footnote{See Einstein to Schwarzschild, 29 December 1915 (Vol. 8, Doc. 176 CPAE) and 9 January 1916 (Vol. 8, Doc. 181 CPAE). To be precise, Einstein was still a week away from what we today call the Einstein field equations when he submitted the Mercury
paper. However, the Schwarzschild metric is a solution both to the Einstein equations and to the field equations Einstein presupposed when he wrote the paper, and his metric approximates the Schwarzschild metric.} It was thus a source of great surprise (and joy) to Einstein when on 22 December 1915 (Vol. 6, Doc. 24 CPAE), less than a month after he had published his Mercury paper, Karl Schwarzschild wrote to Einstein that he had found the exact solution that Einstein's solution to the linearized field equations approximated. \citet{Schwarzschild:1916a} had found the unique spherically symmetric, static and asymptotically flat exact solution to Einstein's vacuum field equations.\footnote{ \cite{Schwarzschild:1916b} would soon deliver an interior extension to the exterior Schwarzschild solution that replaces the singularity of the exterior solution by an incompressible fluid. The matched exterior and interior solutions give a realistic model of the Sun and its exterior gravitational field. For discussion on the interpretation of the (approximative) exterior Schwarzschild solution in Einstein's Mercury paper of 1915, see \citet{Lehmkuhl:2017a}.}

For a mathematician of Hermann Weyl's calibre, the discovery of the Schwarzschild solution must have offered an immediate and clearly formulated challenge. After all, spherical symmetry is a rather special symmetry: it is a special case of axial symmetry. Moreover, while spherical symmetry is a good approximation of most (if not all) isolated astronomical bodies, two-body systems and especially $n$-body systems would be naturally represented by assuming axial symmetry. So Weyl set out to find the generalization of the unique Schwarzschild solution to the class of all axially symmetric static solutions. In \citet{Weyl:1917axial}, Weyl gave the general metric for this class of solutions:\footnote{\citet{EinsteinGrommer:1927} use the notation of \citet{Bach:1922axial}; I follow them in this regard with only a slight change regarding how to express partial derivatives.}

\begin{equation}
	ds^2 = e^{2 \psi} dt^2 - e^{-2 \psi} [r^2 d \theta^2 + e^{2 \gamma} (dr^2 + dz^2)]
	\label{E:Weyl metric}
\end{equation} 

The metric is expressed using what would soon be called `Weyl's canonical coordinates' $(r, z)$. $\theta$ is the azimuth of the meridian plane, $z$ and $r$ are coordinates in this plane, and the condition $r=0$ gives us the $z$-axis as the symmetry axis of rotation. The functions $\psi$ and $\gamma$ depend only on $(r,z)$, and are governed by the equations

\begin{equation}
	\Delta \psi = \frac{1}{r} [\frac{\partial}{\partial z}(r\frac{\partial \psi}{\partial z}) + \frac{\partial}{\partial r}(r\frac{\partial \psi}{\partial r})] = 0
	\label{E:Poisson psi}
\end{equation}

\noindent and

\begin{equation}
	d \gamma = 2 r \frac{\partial \psi}{\partial z} \frac{\partial \psi}{\partial r} dz + r (\frac{\partial^2 \psi}{\partial r^2} - \frac{\partial^2 \psi}{\partial z^2}) dr
	\label{E:dgamma}
\end{equation}

The vacuum Einstein equations, $\R=0$, do indeed reduce to the equations (\ref{E:Poisson psi}) and (\ref{E:dgamma}) for the special case of static axially symmetric fields. This is particularly striking as (\ref{E:Poisson psi}) is effectively the linear Poisson equation, which has lead to $\psi$ being identified with the Newtonian potential. However, this does not mean that the Einstein equations reduce to the Newtonian gravitational equations for the case of static and axially symmetric fields; the equation (\ref{E:dgamma}) encodes the non-Newtonian dynamics of these fields, and in contrast to (\ref{E:Poisson psi}), it exhibits non-linear features.\footnote{The reader will realise that I barely scratch the surface in giving an account of this class of solutions; I have to constrain myself to the aspects that we need to consider in order to decipher the Einstein-Grommer paper. For more details on the Weyl class of solutions see \citet{Stephani:2009}, sec. 20.2 and \citet{GriffithsPodolsky:2009}, chapter 10.} 

So far so good. Now, already in 1917 Weyl turned his attention to a particularly important class of axisymmetric target systems: $n$-body systems. Weyl models $n$ extended and approximately spherical bodies by dust energy-momentum tensors, and gives the metric for this system, and for the special case of a two-body system in particular.  One thing that Weyl immediately discussed was the question as to  whether these solutions had any singularities, and if so how problematic they were. In the following, I will focus on this special case of the axially symmetric field of two bodies, summarise the development of  Weyl's thought on the questions of whether there are singularities in this solution, and as to how they ought to be interpreted. We will see that Einstein \emph{must} have reviewed Weyl's 1917 to 1922 papers on the topic following his correspondence with Rainich, and that in the Einstein-Grommer paper the authors  both draw on and reinterpret Weyl's results. This reinterpretation uncovers how Einstein distinguished between `good and bad singularities'; a distinction that is strongly informed by his picture of the role of matter in GR. 

\subsection{Weyl's vs Einstein's interpretations of the singularities in the two-body solution}

\subsubsection{Weyl stresses}

Already in \citeyear{Weyl:1917axial}, and in more detail in \citet{Weyl:1922axial}, Weyl notes that there are two conditions that the static axisymmetrical metric (\ref{E:Weyl metric}) for the special case of a two-body system has to fulfil in order to be regular (meaning `free of singularities'). He distinguishes between two cases and finds two conditions:

\begin{enumerate}
	\item[i.)] In order for $\g$ to be regular \emph{outside} of the rotation-axis (the $z$-axis), it is sufficient for $\psi$ to be regular.
	\item[ii.)]	In order for $\g$ to be regular \emph{on} the rotation-axis, $\gamma$ has to vanish on the axis.	
\end{enumerate}

Commenting on \citet{Levi-Civita:1918a}, who had criticized Weyl for not having given the most general form of static and axially symmetric metrics, \citet{Weyl:1919axial} explained that while Levi-Civita was interested in all such metrics, he himself was interested in those that could be interpreted as being produced by an axially symmetric distribution of masses. Thus, even though Weyl's class of metrics (\ref{E:Weyl metric}) is a solution to the vacuum equations, Weyl himself wanted to look primarily at those members in the class that could be interpreted as due to sources/regions for which $\T \neq 0$. He then singles out the special case in which the mass density $T^4_4$ is the only component of $\T$ which does not vanish in the source-regions, i.e. he looks at the special case of a distribution of relativistic dust as the source of a static axially symmetric metric. Now comes the part that must have made Einstein perk up, as we will shortly see. Weyl notes that of course a static metric needs to have sources that are at rest with respect to one another. On the other hand, a distribution of dust particles ought to move towards one another under the influence of gravity. Correspondingly, Weyl finds that the only way for the distribution of masses to remain static (and thus to give rise to the metric (\ref{E:Weyl metric})) is for there to be stresses \emph{between} the masses that counteract gravitational attraction. In other words, Weyl finds that the \emph{supposedly matter-free region} between the dust particles contains stresses. Concretely, he finds that in order for the masses to remain static, it needs to be the case that 

\begin{equation}
	T_1^1 + T_2^2 = 0
	\label{E:Weyl stress}
\end{equation} 

along the rotation axis connecting the two particles. Using some of the results of \citet{Levi-Civita:1918a}, Weyl then expresses $T_1^1$ and $T_2^2$ in terms of derivatives of $\psi$ and $\gamma$.

\citet{Weyl:1922axial} further elaborates on the meaning of  condition (\ref{E:Weyl stress}) for the special case of the two-body problem. He argues that in order for two bodies to be at rest with respect to one another, we need stresses in the Meridian plane $(r, z)$: ``a pull in one direction and, orthogonally to the pull, a pressure of the same strength''.\footnote{\citet{Weyl:1922axial},  p.311. As we will see in the next section, the alternative to admitting the stresses (\ref{E:Weyl stress}) is to allow for a conical singularity between the two bodies. This singularity has become known as a `Weyl strut' that holds the two particles apart. } Finally, Weyl argues that condition (\ref{E:Weyl stress}) suffices to fulfil condition ii.) above: the condition makes $\gamma$ vanish on the axis and thus avoids a line singularity along the axis.\footnote{To be precise, Weyl notes that  the stresses given by condition (\ref{E:Weyl stress}) suffice to ``get rid of the worst singularity, a $\gamma$ different from $0$; but they do not suffice to get rid of the singularity completely, for the derivative [$\frac{\partial \gamma}{\partial r}$] on the axis will nevertheless not vanish in general.''} 

%All this must have been very different from what Einstein remembered when he wrote to Rainich of the static and axial symmetric solutions by Weyl and Levi-Civita, claiming that they were capable of representing the exterior field of particles represented as singularities. At least Weyl did not think of his solutions in this way, and actively modeled the particles via non-vanishing mass density, and only avoided a line singularity between the dust particles modeled thus by introducing stresses in the supposedly matter-free region between the particles. 

\subsubsection{Einstein's reinterpretation of Weyl's results}

We know from Einstein's letter to Rainich of 18 April 1926 that at that time he thought of Weyl's two-body solution as ``represent[ing] the case of two electrons at rest (as singularities)''.\footnote{See section \ref{S:Rainich}.} But in his letter of 6 June 1926  he writes that he had been wrong in thinking that representing material particles by singularities implied the possibility that they are at rest with respect to one another; a possibility that Rainich and Einstein had agreed would render the theory empirically inadequate. But why would he have changed his mind after (re)-reading the papers by Weyl, Levi-Civita, and Bach? After all, have we not just seen that Weyl gives a static two-body solution? Why did Einstein not think that what Weyl thought of as a static two-body solution was a \emph{genuine} static two-body solution? Why did he believe, as of 6 June 1926, that  one could instead derive \emph{the motion} of one body from knowing the gravitational field of another body?

Of course, first there is the matter that Weyl introduced non-vanishing $\T$-components in the region between the two bodies that was supposedly free of matter. Having Einstein's reservations regarding $\T$ in mind, there is no doubt that this would have been unacceptable to Einstein. But having singularities between the two bodies would have been equally unacceptable  --- GR was supposed to give an exact description of the spacetime regions outside of material bodies. Einstein needed a way to avoid the line singularity along the rotation axis that did not rely on Weyl's introduction of non-vanishing $\T$-components.\footnote{This was not the first time that Einstein had deemed a particular two-body solution  unsatisfactory. \citet{Trefftz:1922} had published a solution to the vacuum Einstein equations with cosmological constant that he interpreted as a static two-body solution for the special case in which the exterior field of the total system exhibits spherical symmetry. \citet{einstein1922bemerkung} argued that the solution exhibits a singularity in the region exterior to the material bodies, which he judged unacceptable; we see the same pattern as in the present case. For analysis of the solution and the different arguments of Trefftz and Einstein see \citet{Giulini:2015}, which generalizes Trefftz' solution to a solution of the Einstein-Maxwell equations with cosmological constant.}

In section 2 of the Einstein-Grommer paper we see Einstein and Grommer wrestle with the Weyl class of solutions. The equations, notation and mathematical details of the section show clearly that Einstein and Grommer must have plowed through \citet{Weyl:1917axial, Weyl:1919axial}, \citet{Bach:1922axial} and \citet{Weyl:1922axial} with a fine comb. But at some point they depart from Weyl, and their interpretation of the math is entirely different. Most importantly, while Weyl clearly thought that one singularity is just as bad as another, Einstein and Grommer were pickier. Looking at the unfolding of the argument in detail, we will see that energy-momentum tensors and singularities were on a par for Einstein in terms of their limited capabilities of representing matter. 

Einstein and Grommer start with the Weyl metric (\ref{E:Weyl metric}) and the field equations for static and axially symmetric metrics in terms of $\psi$ and $\gamma$, equations (\ref{E:Poisson psi}) and (\ref{E:dgamma}). Like \citet{Weyl:1917axial}, they ask for the conditions that have to be fulfilled to make sure that the metric is regular, and come to Weyl's conditions i.) and ii.) described above. The rest of the argument then focuses on the question of what has to happen, and what kinds of moves are allowed, to fullfil condition ii.), i.e. to make $\gamma = 0$ on the rotation axis to avoid a line singularity along the axis. 

Weyl had ensured that $\gamma = 0$ on the axis by allowing for stresses $T_1^1$ and $T_2^2$ in between material bodies. But Einstein and Grommer were working within the vacuum approach, where $\T$ was supposed to vanish everywhere; so this option was not open to them.

Here is what they did instead. Einstein and Grommer first looked at a special member of the class of Weyl metrics (\ref{E:Weyl metric}), namely the one for which

\begin{equation}
	\psi_1 = - \frac{m}{r^2 + z^2}
	\label{E:psi Curzon}
\end{equation}

This is the Newtonian potential of a point mass.\footnote{Remember that Weyl had aimed to find a solution describing the gravitational field of  two extended material bodies described by a dust energy-momentum tensor. Einstein and Grommer, instead, start out by looking for the GR-counterpart of the vacuum solution of the Poisson equation that represents the gravitational field of a point particle.} If we use it to derive $\gamma$ from (\ref{E:dgamma}) we get 

\begin{equation}
	\gamma_1 = \frac{m^2}{2} \frac{r^2}{r^2 + z^2} \quad .
	\label{E:gamma Curzon}
\end{equation}

Plugging both $\psi_1$ and $\gamma_1$ into the Weyl metric (\ref{E:Weyl metric}), we obtain the so-called Curzon-Chazy solution, a particular Weyl metric that is often interpreted as the static axially symmetric field of a single particle.\footnote{The solution is named after \citet{Curzon:1925a} and \citet{Chazy:1924a}, both of whom actually presented a solution that is a superposition of two now so-called Curzon particles as produced by equation (\ref{E:psi Curzon}). For mathematical details see \citet{GriffithsPodolsky:2009}, sec. 10.5.  There is no evidence that Einstein and Grommer knew of Curzon's and Chazy's papers. However, in 1936 Ludwig Silberstein independently discovered the Curzon-Chazy two-body solution, and a heated debate between Einstein and Silberstein on the interpretation of the solution ensued, in which Einstein and his collaborator Nathan Rosen argued that it is a non-physical solution because of a conical singularity between the two bodies. There is no sign of Einstein remembering that he had already struggled with this solution and these questions in 1927. For details of the Einstein-Silberstein debate see \citet{Havas1993}.}

The Curzon-Chazy metric has a curvature singularity at $(r = 0, z = 0)$.\footnote{It is a special kind of curvature singularity that has a directional dependence; i.e., things are different depending on the direction from which one approaches the singularity.  The singularity also has the structure of a ring, which leads \citet{GriffithsPodolsky:2009}, section 10.5, to question whether it really should be interpreted as the GR-counterpart of the Newtonian gravitational field of a point particle, despite the fact that the latter is the starting point for deriving it.} Einstein and Grommer clearly interpret this singularity as the `location' of the particle that gives rise to the  metric. It is also clear that they have no problem with admitting and accepting as non-problematic \emph{this} singularity. And yet, they spend the rest of the section trying to avoid another singularity: the one that threatens to occur on the rotation axis if Weyl's condition ii.), the vanishing of $\gamma$ along the axis, is not fulfilled.\footnote{I will follow the modern literature in the following and speak of a Curzon particle as a particle modelled by the exterior field given by equations (\ref{E:psi Curzon}) and (\ref{E:gamma Curzon}) plugged into equation (\ref{E:Weyl metric}). This terminology is less confusing than speaking of, as Einstein and Grommer do,  `the singularity' when what is referred to is a Curzon metric with a curvature singularity at its center; especially given that there is another singularity threatening that is not interpreted  as corresponding to a material particle.} 

First, Einstein and Grommer note that as things stand it follows directly from equation (\ref{E:gamma Curzon}) that $\gamma$ vanishes; if there is only one particle then there is no danger of an additional singularity. But now Einstein and Grommer proceed to the question of what happens when the  particle is subject to an external gravitational field. Thus, they turn Weyl's two-body problem into the problem of one body subject to an external field, thereby creating a clear relationship between the Weyl class of solutions and the question whether we can derive the equations of motion of a body/particle from the external field (equations) that the body is subject to. 

Einstein and Grommer find the metric of a particle $\psi_1$ subject to an external field $\hat{\psi}$ with the ansatz

\begin{equation}
	\psi_{total} = \psi_1 + \hat{\psi} \quad . 
	\label{E:psi total}
\end{equation}

Here, $\psi_1$ is interpreted as corresponding to the exterior metric produced by the particle and characterizing the particle in turn, $\hat{\psi}$ the exterior field that it is subject to. They assume that $\hat{\psi}$, just like $\psi_1$, is a function of only $(r, z)$, and that it is regular in the neighbourhood of $r = z = 0$, i.e. in the neighbourhood of the singularity corresponding to the particle. The presumed axial symmetry of the total system then allows them to assume that  $\hat{\psi}$ has the form

\begin{equation}
	\hat{\psi} = \alpha_0 + \alpha_1 z + G
	\label{E:external field}
\end{equation} 
 
where $G$ includes the terms of second and higher order in $r$ and $z$, and $\alpha_0$ and $\alpha_1$ are constants.

Calculating $\gamma$ by plugging $\psi_{total}$ into equation $(\ref{E:dgamma})$, Einstein and Grommer find that $\gamma$ is constant along the positive $z$-axis (the rotation axis), and that one can thus set $\gamma = 0$ for $(r = 0, z > 0)$, as demanded by Weyl's condition ii.). What remains to be found is a condition that makes $\gamma = 0$ for $(r = 0, z < 0)$ also, in order for there to be no line singularity along the rotation axis $z$.

Einstein and Grommer now argue that $\gamma = 0$ for $(r = 0, z < 0)$ if and only if the line integral of $d \gamma$ over a closed circle around the singularity at $(r = 0, z = 0)$ vanishes. This, in turn, is only possible if $\alpha_1$ in equation (\ref{E:external field}) is zero. In sum, Einstein and Grommer show that the only way to avoid a line singularity along the $z$-axis is given by the following chain of conditions:

\begin{equation}
	\mbox{No line singularity along z-axis} \iff \gamma = 0 \quad \mbox{when} \quad r \rightarrow 0 \iff \oint_{r \rightarrow 0} d \gamma = 0 \iff \alpha_1 = 0
\end{equation}

The final condition, that $\alpha_1$ has to be zero, is interpreted by Einstein and Grommer as the condition that the external field $\hat{\psi}$ vanishes at $(r =0 , z = 0)$.

% XXX Maybe they should have said that the flow of gamma through the surface integral around the particle vanishes. That's why they think of this as the equilibrium condition generalized in the next section: it's about the flow of the gravitational field \gamma through the surface surrounding the particle!! 

The only alternative is to admit a line singularity along the $z$-axis. In the subsequent literature on the two-body solution, this singularity has been called a `Weyl strut': a singular strut that keeps the two bodies apart.\footnote{See \citet{GriffithsPodolsky:2009}, sec. 10.6 for details on this singularity.} But Einstein wanted to avoid this for reasons discussed in the next section. And the only way to achieve this, Einstein and Grommer argued, was for the external gravitational field to vanish at the location of the particle.\footnote{Einstein and Grommer note without further comment that their argument does not put any constraints on  $G$ of equation (\ref{E:external field}). However, in order for the external field $\hat{\psi}$ to vanish given that $\alpha_1$ vanishes,  $G$ has to vanish as well. Given that  $G$ was to summarize the terms of second and higher order in $r$ and $z$,  Einstein and Grommer might assume that $G$ can be neglected. But this would only be justified in the linearized approximation of the field equations, which Einstein and Grommer only introduce way later in the paper, in section 3. As it stands,  leaving $G$ unconstrained arguably is a liability for the conclusion they want to draw.}   

Thus, Einstein  and Grommer effectively conclude that in the full, non-linear theory, there is no \emph{physical} solution of a particle at rest but subject to an external gravitational field. In turn, they say, it follows from the gravitational field equations that a particle cannot be at rest when subject to a gravitational field. For if it \emph{were} at rest, then there would also have to be a singularity in matter-free space.  

All this is of course very different from Newtonian mechanics, Maxwellian electrodynamics \emph{and} indeed the linearized Einstein equations. In each of these cases the field equations \emph{by themselves} allow for a particle at rest while subject to an external field, or for two particles to be at rest with respect to one another despite exerting forces on one another. But in GR, the field equations by themselves predict whether a particle moves when subject to an external field; they predict \emph{that} it will move. From here it is only a small step to expect the field equations to determine \emph{how} the particle will move.\footnote{This last step is the concern of the rest of the Einstein-Grommer paper. I have summarized it in section 3.2 of \citet{Lehmkuhl:2017a}, but future work will be required to show whether the argument stands up to scrutiny. Note also that one might be tempted to say that the results of Einstein and Grommer do not actually depend on assuming the vacuum Einstein equations, but only on assuming that the metric of spacetime is a particular Weyl metric. Likewise, one might argue that \citet{Einsten:1915h} did not derive the perihelion of Mercury from the field equations but from assuming that the metric around the Sun is an approximation of the Schwarzschild metric; which solves many field equations in addition to  $\R = 0$. But while the argument goes through for Mercury, it does not for the case at hand. For the Einstein equations for axially symmetric fields reduce to equations (\ref{E:Poisson psi}) and (\ref{E:dgamma}), and assuming these equations played a vital role in the argument above.}

But now let us take a step back. The whole argument depends on the assumption that it is acceptable to have a curvature singularity at $(r = 0, z = 0)$, whereas it is not acceptable to have a line singularity along the axis of rotation (the $z$-axis). Why is that? Where does this distinction between good and bad singularities come from?

\section[$T_{\mu \nu}$ vs singularities]{The representational capabilities of energy-momentum tensors vs. singularities}

The difference between the two singularities of the Weyl metrics discussed by Einstein and Grommer is \emph{not} about intrinsic vs. coordinate singularities. Both of them are intrinsic singularities, if they exist. No, the difference is that the curvature singularity at $(r = 0, z = 0)$ stands in for the material particle, while the line singularity, if it were to occur, would be in `free space', the part of space free of matter. 

As we have seen in section \ref{S:T}, Einstein believed that GR got the laws of pure gravitational fields, i.e. spacetime regions free of matter, essentially right. In this domain, GR is supposed to be a fundamental theory, and thus one free of singularities. In spacetime regions in which matter is present, Einstein believed, GR is only capable of giving a phenomenological picture frame anyhow, a frame that needs to be filled by a proper picture of matter. Thus, in these spacetime regions all GR can provide is a placeholder for a theory of matter not yet delivered, a docking station for the correct theory of matter not yet found. Hence, it does not matter whether this placeholder-role is fulfilled by an energy-momentum tensor or by a curvature singularity; neither of them is supposed to be taken as a serious representation of matter anyway.\footnote{See \citet{EarmanEisenstaedt:1999}, section 7, for Einstein quotes from the 1930s and 1940s that provide further evidence for this interpretation.} 

In 1922, Einstein had spoken of representing material particles by singularities as a pseudo-solution. In the course of his correspondence with Rainich, he had come to believe that this pseudo-solution might give us at least a hint of how to get to the real prize: a complete field theory of gravity \emph{and} matter (and electromagnetism). As we saw in section \ref{S:T}, for Einstein such a theory had to, first and foremost, give an account of the discretization of charge and energy, which Einstein considered the essence of the quantum nature of matter. 

I believe this is what Einstein  had in mind when he wrote again to Paul Ehrenfest, only five days after presenting the Einstein-Grommer paper to the Prussian Academy on 6 January 1927. He wrote:\footnote{Vol. 15, Doc. 450 CPAE. The catch Einstein refers to is  likely (as alluded to in the conclusion of the paper) that the result only applies to electrically neutral particles. In a follow-up paper ten months later, Einstein would aim to derive the equations of motion of a charged particle from the Einstein-Maxwell equations.}

\begin{quote}
The problem of motion turned out nicely, even if a small catch remains. Anyway, it is interesting that field equations can determine the motion of singularities. I even think that this will one day determine the development of quantum theory, but the path to this is still in the dark.
\end{quote}	

Remember that 5 months before (and 2 months after his last letter to Rainich) Einstein had written to Michele Besso that ``[i]t is questionable whether the equation $R_{ik} - \frac{1}{2} g_{ik} R = T_{ik}$ has any reality left within it in the face of quanta. I vigorously doubt it. In contrast, the left-hand side of the equation surely contains a deeper truth.''\footnote{See section \ref{S:T}.} 

He went on directly after this:\footnote{Einstein to Michele Besso, 11 August 1926 (Vol. 15, Doc. 348 CPAE).}

\begin{quote}
If the equation $R_{ik} = 0$ really does determine the behaviour of the singularities, then the resulting law would be founded more deeply than by the aforementioned equation, which is not of a piece, and only phenomenologically founded. 
\end{quote}	

Again, the complaint about the merely phenomenological nature of the full Einstein equations was about $\T$ not being able to account for the quantum nature of matter. Thus, for Einstein there was a difference between using energy-momentum tensors and using singularities to stand in for matter after all. The former were helpless in the face of quanta. With regard to the latter, Einstein saw at least a hope that their behaviour, as derived from the fundamental laws of the pure gravitational field, would give us an inkling of where to go next.

\section{Summary}
\label{S:Lessons}
In this paper, I have looked at the genesis of Einstein's first paper on the problem of motion, the idea that in GR the equations of motion of material particles might be derivable from the field equations themselves. We have seen that the key to Kennefick's puzzle, the question of why Einstein never ever commented on this possibility between 1913 and 1927, lies in the Einstein-Rainich correspondence.

Along the way, we have found a hitherto unappreciated facet of Einstein's own interpretation of GR. General Relativity, as interpreted by Einstein, is not a fully fundamental theory. It is not a fully effective theory either. It is a hybrid theory, fundamental in its treatment of purely gravitational fields in regions of spacetime in which nothing else is present, and effective/phenomenological in regions of spacetime in which matter is present. Contrary to the widespread idea that Einstein was more and more opposed to quantum theory, he strongly believed that an adequate theory of matter would have to do justice, first and foremost, to its quantum aspects. Of course, for Einstein this meant to account for the discreteness features of quantum particles; Einstein \emph{was} opposed to regarding the new quantum mechanics (matrix and wave mechanics) as a final and complete description of reality. But then, he also did not regard GR as a final and complete description of reality. 

From early on, Einstein saw GR as a theory of the pure gravitational field, extremely wary of it putting any constraints on what kinds of matter give rise to the gravitational field. Even in the Einstein-Grommer paper he clearly forbids singularities \emph{outside} of material particles (where the theory is supposed to give an adequate and deterministic representation of gravitational fields) but has no problem with them appearing \emph{inside of} material systems, where the theory can provide at best phenomenological placeholders for a future `proper' theory of matter anyhow. Thus, for Einstein energy-momentum tensors as alleged representatives of material systems were on a par with singularities: both were only placeholders for a proper (quantum) theory of matter. 

\emph{This} is why Einstein was so adamant in his search for a unified field theory. Yes, he liked the idea that there is only one fundamental field. But more importantly, no one was as aware of the representational shortcomings of GR, as far as matter was concerned, as Einstein. This fuelled  his search for a unified field theory, while the search for the latter and what he expected of a complete theory both strongly informed his interpretation of GR itself. 

Finally, we now understand how Einstein could have allowed for singularities to account for matter in GR, yet be adamant that \emph{no} singularities were allowed to occur in the sought-after unified field theory. It was the hybrid character of GR that allowed for this double standard: it was exactly because it was not supposed to be an adequate theory of matter that it was acceptable to allow for singularities as  place-holders of matter. But it was \emph{not} acceptable to allow for singularities in the domain about which GR was supposed to be fundamental, correct: regions of spacetime with only gravitational fields, free of matter. Accordingly, a unified field theory of gravity \emph{and} matter would have to live up to these latter, stricter standards: no singularities anywhere. 

% GR was a stepping stone for Einstein, from the very beginning. We are still standing perched on our toes, looking up from that same stone. But at least now we know a bit better in which direction Einstein was looking, and we can thus decide better if that's where the promised land of a complete field theot  

\section{Acknowledgements}

I would like to thank my colleagues at Caltech and at the Einstein Papers Project for many discussions about the problem of motion and the Einstein-Grommer approach in particular. I would like to thank especially Patrick D\"urr, Domenico Giulini, Anne Kox, James Read and Jim Weatherall for carefully reading earlier versions of this paper, and for the extremely helpful comments they gave me. Finally, I would like to thank Harvey Brown for many years of inspiration and dialogue, for being a mentor and a friend --- and for making me puzzle about geodesic motion in the first place. 

% Sketch the POM approach based on solutions to the Vacuum equations and characterzing bodies by their exterior gravitational field. 

%\section{Summary}

% Make clear that the EInstein-Grommer results do not
% use the two letters to Ehrenfest I quoted in last exec meeting in last section.

% Cut from PSA Paper (for printed version): \footnote{If I had given more historical details, I could have, I believe, shown that Einstein himself saw the occurence of a singularity in the inner metric in exactly this way. This exegetical argument would have started with evidence that, from early on, he saw GR as a theory of the pure gravitational field without any constraints on what kinds of matter give rise to the gravitational field. Furthermore, I would have argued that even in the Einstein-Grommer paper he clearly forbids singularities \emph{outside} of material particles (where the theory is supposed to give an adequate and deterministic representation of gravitational fields) but has no problem with them appearing \emph{inside of} material systems, where the theory can provide at best phenomenological placeholders for a future `proper' theory of matter anyhow. Thus, for Einstein energy-momentum tensors as alleged representatives of material systems were on a par with singularities:  both were only placeholders for a proper theory of matter. }  

\end{document}